\documentstyle[aps,prl,twocolumn]{revtex}
\include{psfig}

\begin{document}
\draft 

\title{Configurational Entropy and Diffusivity of Supercooled Water}

\author{Antonio Scala$^{1}$, Francis W.~Starr$^{1,}$\thanks{Current
Address: Polymers Division and Center for Theoretical and Computational
Materials Science, National Institute of Standards and Technology,
Gaithersburg, Maryland 20899}, Emilia La Nave$^{1}$, \\ Francesco
Sciortino$^{2}$ and H.~Eugene Stanley$^{1}$}

\address{$^{1}$ Center for Polymer Studies, Center for Computational
Science, and Department of Physics, \\ Boston University, Boston, MA
02215 USA}

\address{$^{2}$ Dipartimento di Fisica e Istituto Nazionale per la
Fisica della Materia, \\ Universit\'a di Roma {\it La Sapienza}, P.le
Aldo Moro 2, I-00185, Roma, Italy }

\date{\today} 
 
\maketitle
 
\begin{abstract}
We calculate the configurational entropy $S_{\mbox{\scriptsize conf}}$
for the SPC/E model of water for state points covering a large region of
the $(T,\rho)$ plane.  We find that (i) the $(T,\rho)$ dependence of
$S_{\mbox{\scriptsize conf}}$ correlates with the
diffusion constant and (ii) that the line of maxima in $S_{\mbox{\scriptsize
conf}}$ tracks the line of density maxima. 
Our simulation data indicate that the
dynamics are strongly 
influenced by $S_{\mbox{\scriptsize conf}}$ even above the
mode-coupling temperature $T_{\mbox{\scriptsize MCT}}(\rho)$.
\end{abstract}
\bigskip

Computer simulation studies are contributing to the understanding of the
slow dynamical processes in simple and molecular liquids on approaching
the glass transition\cite{kob-review}.  In particular, the space, time
and temperature dependence of many dynamical quantities have been
calculated and compared with the predictions of the mode-coupling theory
(MCT)~\cite{mct}, which provides a description of the dynamics in weakly
supercooled states~\cite{WSSnote} of many atomic and molecular liquids,
including the SPC/E model of water studied here~\cite{francesco,shss}.

Simulations have also been used to investigate the ``thermodynamic
approach'' to the glass transition~\cite{adam-gibbs,parisi,speedy},
which envisages a relation between diffusion constant $D$ and
configurational entropy $S_{\mbox{\scriptsize conf}}$ by relating the
dynamics of liquids at low $T$ to the system's exploration of its
configuration space.  The properties of the liquid are dominated by the
basins of attraction of local potential energy minima; the liquid
experiences vibrations localized around a basin and rearranges via
relatively infrequent inter-basin jumps~\cite{goldstein}.  Motivated by
the separation of time scales, the total entropy $S$ may be separated
into two parts: (i) an intra-basin contribution $S_{\mbox{\scriptsize
vib}}$, which measures the vibrational entropy of a system constrained
to reside within a basin, and (ii) a configurational contribution,
$S_{\mbox{\scriptsize conf}}$, quantifying the multiplicity of
basins~\cite{COMPLEXnote}.  Explicit calculation of
$S_{\mbox{\scriptsize conf}}$ has been performed for hard
spheres~\cite{speedy}, soft spheres\cite{parisi}, Lennard-Jones
systems~\cite{barbara,skt99} and  tetravalent network glasses\cite{robinpablo}
with the aim of evaluating the Kauzmann
temperature $T_K$, at which $S_{\mbox{\scriptsize conf}}$ appears to
vanish~\cite{kauzmann}.

\begin{figure}
\centerline{\mbox{\psfig{figure=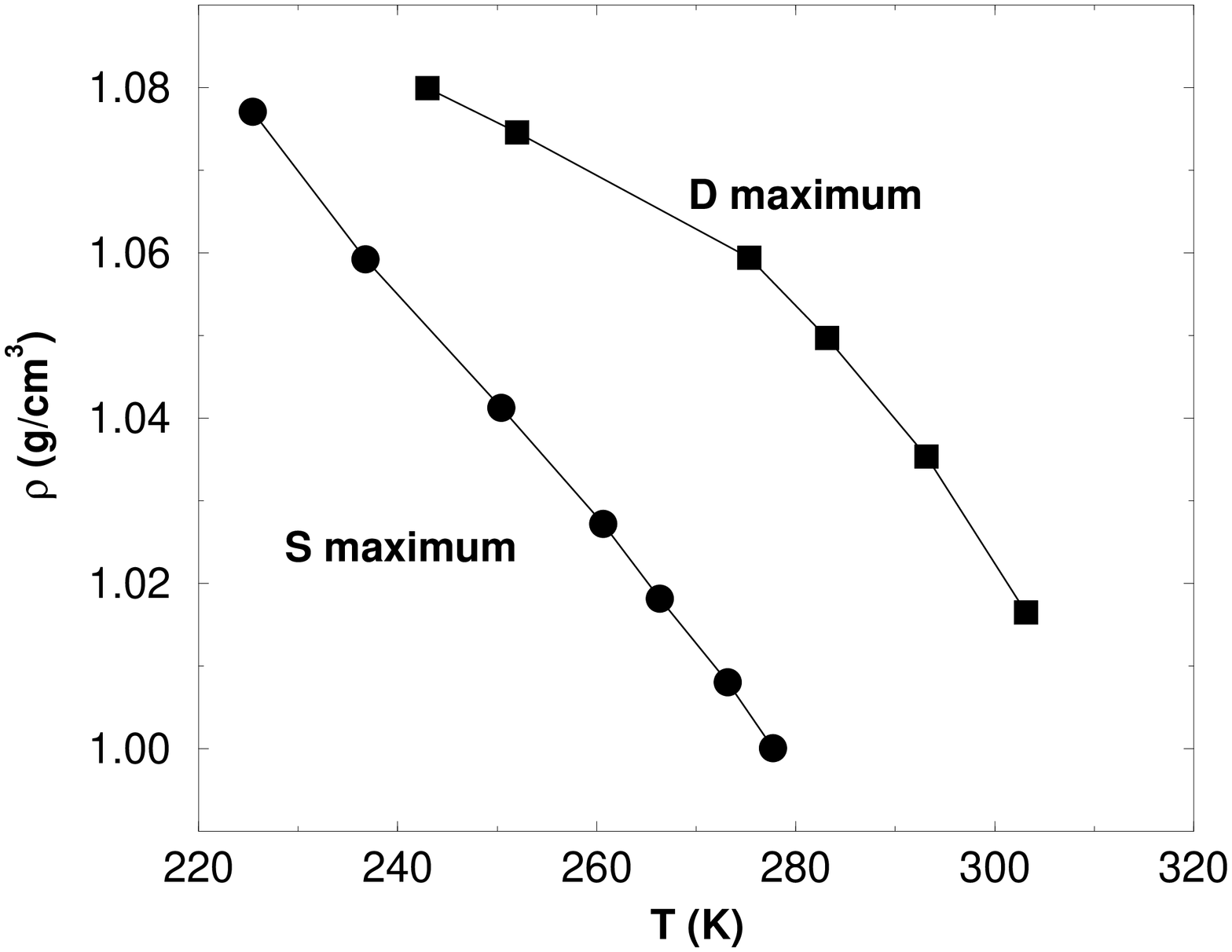,width=8.0cm,angle=-90}}}
\centerline{\mbox{\psfig{figure=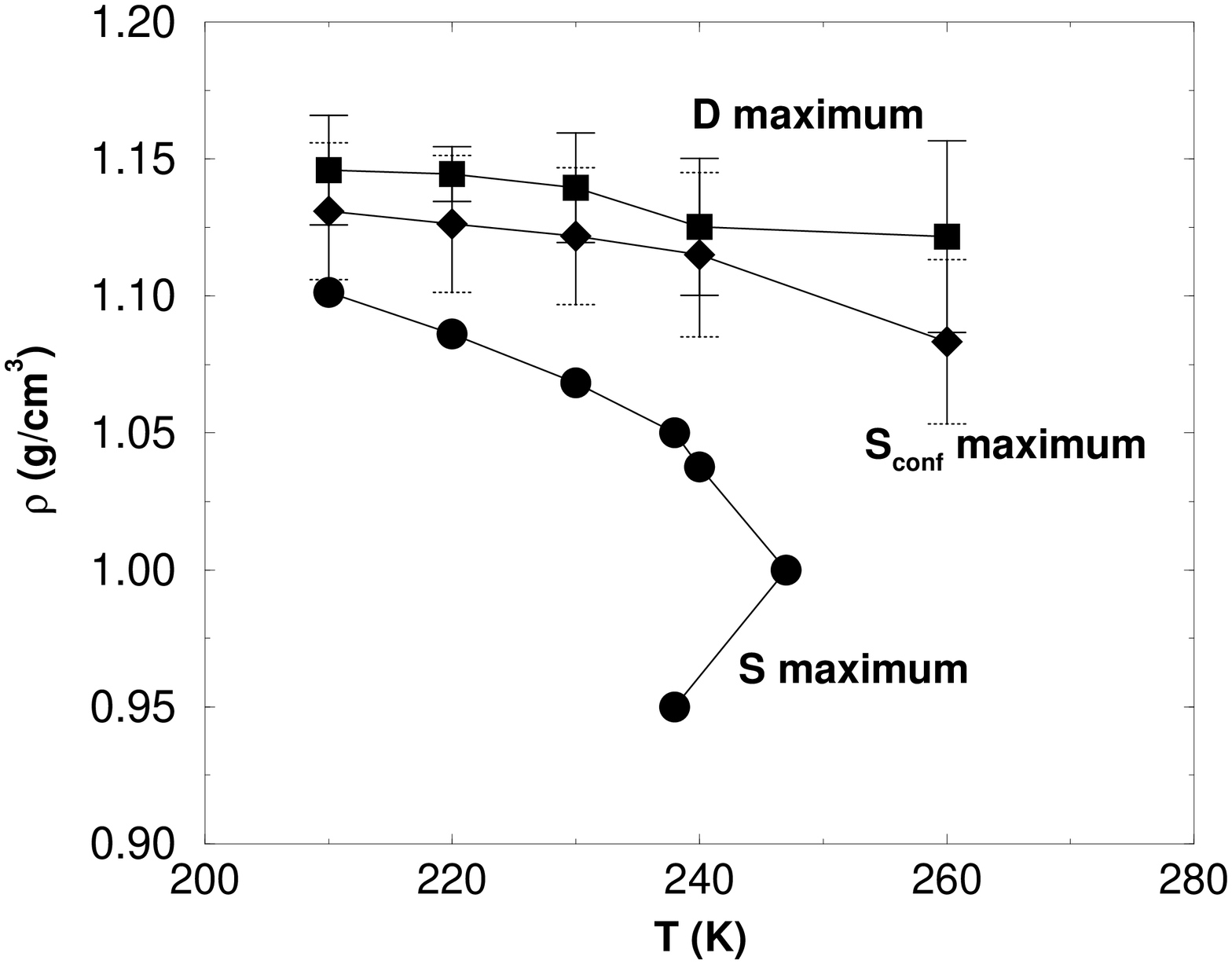,width=8.0cm,angle=-90}}}
\caption{(a) Line of maxima of the entropy $S$ and of 
the diffusion constant $D$ from experimental data on
water~\protect\cite{prielmeier}.  (b) Lines of entropy maxima $S$,
configurational entropy maxima $S_{\mbox{\scriptsize conf}}$ and
diffusion constant maxima $D$ (from Ref.~\protect\cite{shss}) for the
SPC/E model. }
\label{fig:waterSDmax}
\end{figure}

One of the most studied models of molecular liquids is the SPC/E
potential, designed to mimic the behavior of water~\cite{spce}.  The
dynamical properties of this model have been studied in detail in the
weakly supercooled regime and have been shown to be consistent with the
predictions of MCT~\cite{francesco,shss}. The SPC/E potential is of
particular interest for testing theories of the supercooled-liquid
dynamics, since, as observed experimentally for water, along isotherms
$D$ has a maximum as function of the pressure $P$ or of the density
$\rho$~\cite{prielmeier}.  This maximum becomes more pronounced upon
cooling.  Furthermore, the line of isobaric density maxima, which
corresponds to the line of isothermal entropy maxima from the Maxwell
relation $(\partial V/ \partial T)_P = - (\partial S / \partial P)_T
$, appears strongly correlated to the line of $D$ maxima
[Fig.~\ref{fig:waterSDmax}(a)], providing motivation to test possible
relationships between $S_{\mbox{\scriptsize conf}}$ and $D$.
A test of such relationship, based on the analysis of experimental data at ambient pressure, 
was performed 
by Angell and coworkers in 1976\cite{austen}.

\begin{figure}
\centerline{\mbox{\psfig{figure=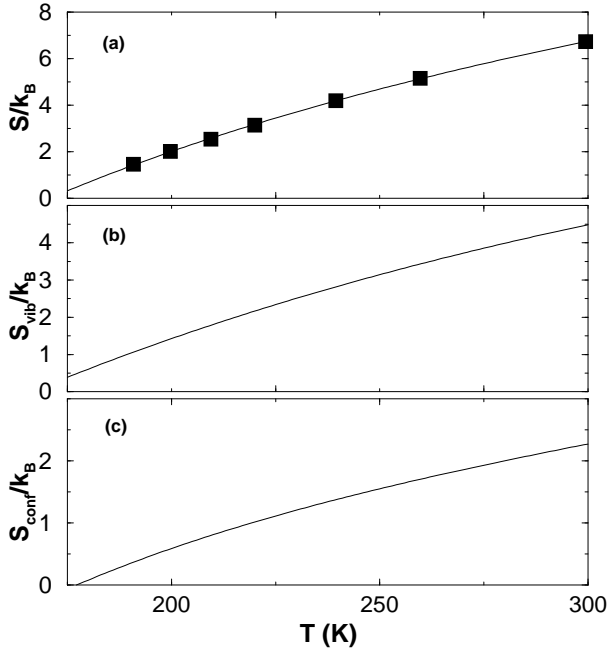,width=8.0cm,angle=-90}}}
\caption{Calculated values for the $\rho = 1.0$~g/cm$^3$ isochore of:
(a) $S$ determined from simulation (filled $\Box$).  The line is
a fit used to extrapolate $S$ to lower $T$~\protect\cite{T35-note}.
(b)$S_{\mbox{\scriptsize vib}}$.  (c) $S_{\mbox{\scriptsize conf}}$,
obtained by methods described in the text.
}
\label{fig:entro}
\end{figure}

Here we calculate the entropies $S$, $S_{\mbox{\scriptsize vib}}$ and
$S_{\mbox{\scriptsize conf}}$ for the SPC/E potential for state points
covering a large region of the $(T,\rho)$ phase diagram.  We then
compare the behavior of $S_{\mbox{\scriptsize conf}}$ with $D$. We also
estimate $T_K$ and consider its relation with the MCT temperature
$T_{\mbox{\scriptsize MCT}}$.  We first calculate $S$ for a reference
point ($T=1000$~K and $\rho=1.0$~g/cm$^3$)\cite{frenkel}, and then
calculate the entropy as a function of $(T,\rho)$ via thermodynamic
integration~\cite{CYCLEnote} using the state points simulated in
Ref.~\cite{shss}, as well as new simulations extending to higher
temperatures~\cite{simulation-note}.  We show an example of $S$ along
the $\rho = 1.0$~g/cm$^3$ isochore in Fig.\ref{fig:entro}(a).

We calculate $S_{\mbox{\scriptsize vib}}$, i.e. the entropy of the
liquid constrained in a typical basin, using the properties of the
basins visited in equilibrium by the liquid.  For each ($T$, $\rho$), we
calculate, using the conjugate gradient algorithm~\cite{num-recipes},
the corresponding local minima -- called inherent structures (IS)
\cite{stillinger} --
for 100 configurations~\cite{CORRELnote}.  We estimate
$S_{\mbox{\scriptsize vib}}$ by adding anharmonic corrections to
the harmonic contribution $S_{\mbox{\scriptsize harm}}$\cite{pohorille}
\noindent 
\begin{equation}
S_{\mbox{\scriptsize harm}} = 
k_B \sum^{6N-3}_{i=1} \left[ \ln\frac{k_B T}{\hbar \omega_i} - 1 \right],
\label{eq:sharm}
\end{equation}
where $\omega_i$ are the normal-mode frequencies of the IS determined
form the Hessian matrix~\cite{landau}.  Ref.~\cite{skt99} found that the
harmonic approximation for a binary Lennard-Jones mixture is a valid
estimate of $S_{\mbox{\scriptsize vib}}$ for temperatures around
$T_{\mbox{\scriptsize MCT}}$.  However, in the case of the SPC/E
potential, we find that there are significant anharmonicities in the
basins. This can be seen from the fact that, if the system were purely
harmonic, the energy $E$ should equal the energy of a minimum
$E_{\mbox{\scriptsize conf}}$ plus the contribution
$E_{\mbox{\scriptsize harm}}=(6N-3) k_B T/2$ of the harmonic solid
approximation.  In contrast with the binary mixture Lennard-Jones case,
we find the vibrational contribution $E_{\mbox{\scriptsize vib}} \equiv
E-E_{\mbox{\scriptsize conf}}$ to be roughly 10\% larger than the
harmonic approximation, even at the lowest temperatures studied.  We
then estimate the anharmonic contributions to $S_{\mbox{\scriptsize
vib}}$ by heating the IS at constant volume and measuring the deviation
of $E_{\mbox{\scriptsize vib}}$ from the harmonic
approximation~\cite{heating-note}.  For each collection of basins
corresponding to a particular $(T,\rho)$ state point, we calculate
$E_{\mbox{\scriptsize vib}}$ in the range $T=0-200$~K and fit our
results using the approximation
\begin{equation}
E_{\mbox{\scriptsize vib}} = E_{\mbox{\scriptsize harm}} + a T^2 + b T^3.
\end{equation}
\noindent By integrating the relation $dS_{\mbox{\scriptsize vib}} =
dE_{\mbox{\scriptsize vib}}/T$, we find
\begin{equation}
S_{\mbox{\scriptsize vib}} = S_{\mbox{\scriptsize harm}} + 2 a T + \frac{3}{2} b T^2
\label{eq:svib}
\end{equation}
We use Eq.~(\ref{eq:svib}) to extrapolate $S_{\mbox{\scriptsize vib}}$
to higher temperatures.

\begin{figure}
\centerline{\psfig{figure=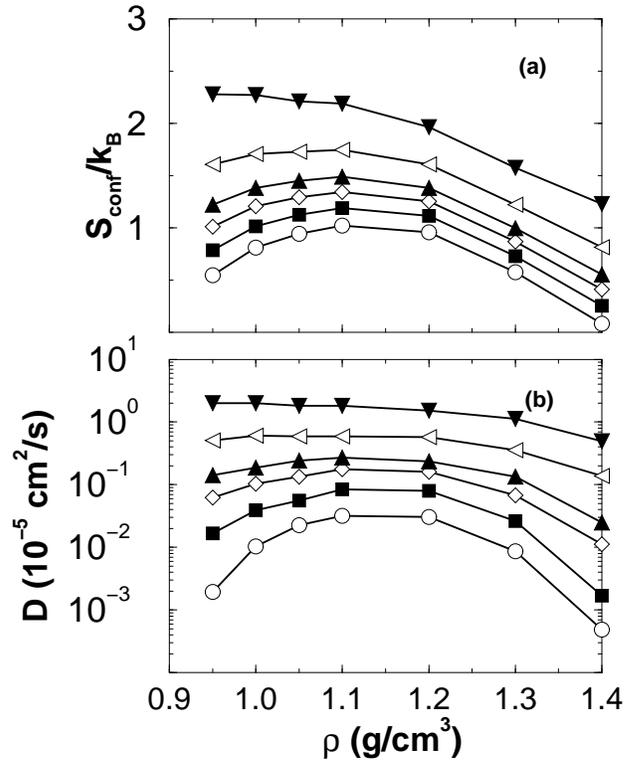,width=8.0cm,angle=-90}}
\caption{(a) The configurational entropy $S_{\mbox{\scriptsize conf}}$
along isothermal paths; from top to bottom $T=300$, $T=260$, 
$T=240$, $T=230$, $T=220$, $T=210$ $K$.
(b) Diffusion constant $D$ along the same
isotherms.  Notice the close correspondence of the maxima of
$S_{\mbox{\scriptsize conf}}$ and $D$.}
\label{fig:S-conf}
\end{figure}

\begin{figure}
\centerline{\psfig{figure=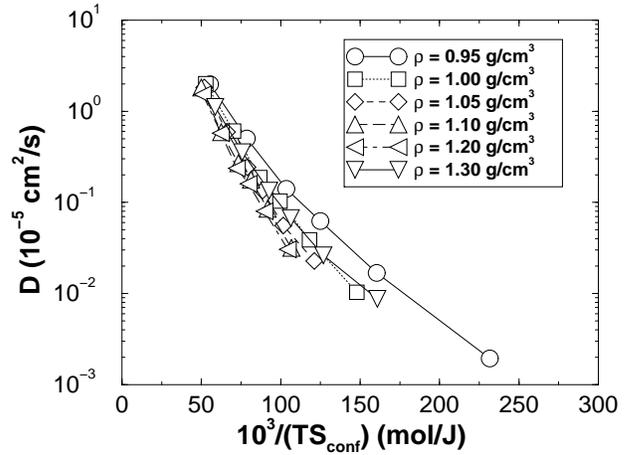,width=8.0cm,angle=0}}
\caption{Diffusion constant $D$ versus $(TS_{\mbox{\scriptsize conf}})^{-1}$
for all studied isochores.}
\label{fig:D-TS}
\end{figure}

>From the knowledge of $S_{\mbox{\scriptsize vib}}$, we calculate
$S_{\mbox{\scriptsize conf}} = S - S_{\mbox{\scriptsize vib}}$.  As an
example of the entire procedure, we show in Figs.~\ref{fig:entro}(a) and
\ref{fig:entro}(b) the $T$ dependence of $S_{\mbox{\scriptsize vib}}$
and $S_{\mbox{\scriptsize conf}}$ for the $\rho=1.0$~g/cm$^3$ isochore.
In Fig.~\ref{fig:S-conf}(a), we show $S_{\mbox{\scriptsize conf}}$ as a
function of $\rho$ for several $T$, and in Fig.~\ref{fig:S-conf}(b) we
show the behavior of $D$ along the same isotherms. Fig.~\ref{fig:D-TS}
shows that at high densities the behavior of $ln(D)$ is nearly linear when
plotted as a function of $(TS_{\mbox{\scriptsize conf}})^{-1}$, as
proposed by Adam and Gibbs\cite{adam-gibbs} over 30 years ago.

Fig.~\ref{fig:waterSDmax}(b) shows the lines of maxima for $D$ and
$S_{\mbox{\scriptsize conf}}$ in the region $T \le 260$~K, where we have
been able to clearly detect a maximum.  To highlight the differences
between $S$ and $S_{\mbox{\scriptsize conf}}$, we show also the line at
which $S(\rho,T)$ has a maximum.  Fig.~\ref{fig:waterSDmax}(b) shows
that the lines of maxima in $D$ and in $S_{\mbox{\scriptsize conf}}$
track each other within the uncertainties of our calculations.

Note that the density where $D$ has a maximum ($\rho \approx
1.15$~g/cm$^3$), as well as the density where the maximum in
$S_{\mbox{\scriptsize conf}}$ occurs, depends weakly on $T$. One
possible explanation is that, for the values of $\rho$ where the maxima
in $D$ and $S_{\mbox{\scriptsize conf}}$ occur, two density-dependent
(and primarily $T$ independent) mechanisms balance.  Increasing $\rho$
in water from the ``ideal'' tetrahedral density (i.e., that of ice Ih,
about 0.92~g/cm$^3$) leads to the progressive destruction of the
hydrogen-bond network, and hence {\it increases} the number of minima
(and therefore S$_{\mbox{\scriptsize conf}}$), since there are more
configurations corresponding to a disordered tetrahedral network.
However, at large enough density, core repulsion begins to dominate the
liquid properties, as expected for ``typical'' liquids.  In such a case,
one expects that increasing $\rho$ {\it decreases} the number of minima
in the potential energy landscape, as the system becomes more densely
packed, and hence fewer configurations are possible.

We observe that the close connection between $D$ and
$S_{\mbox{\scriptsize conf}}$ shown in Fig.~\ref{fig:S-conf} occurs in
the same region where the dynamics of SPC/E water can be rather well
described by MCT, suggesting that MCT may be able to capture the
reduction of the mobility due to entropic effects.  Moreover, in the
same region of the ($T$, $\rho$) plane, $D$ also correlates well with
the number of directions in configuration space connecting different
basins\cite{emilia}.  This suggests the possibility of a statistical
relation between the number of minima and their
connectivity~\cite{angelani}.

\begin{figure}
\centerline{\psfig{figure=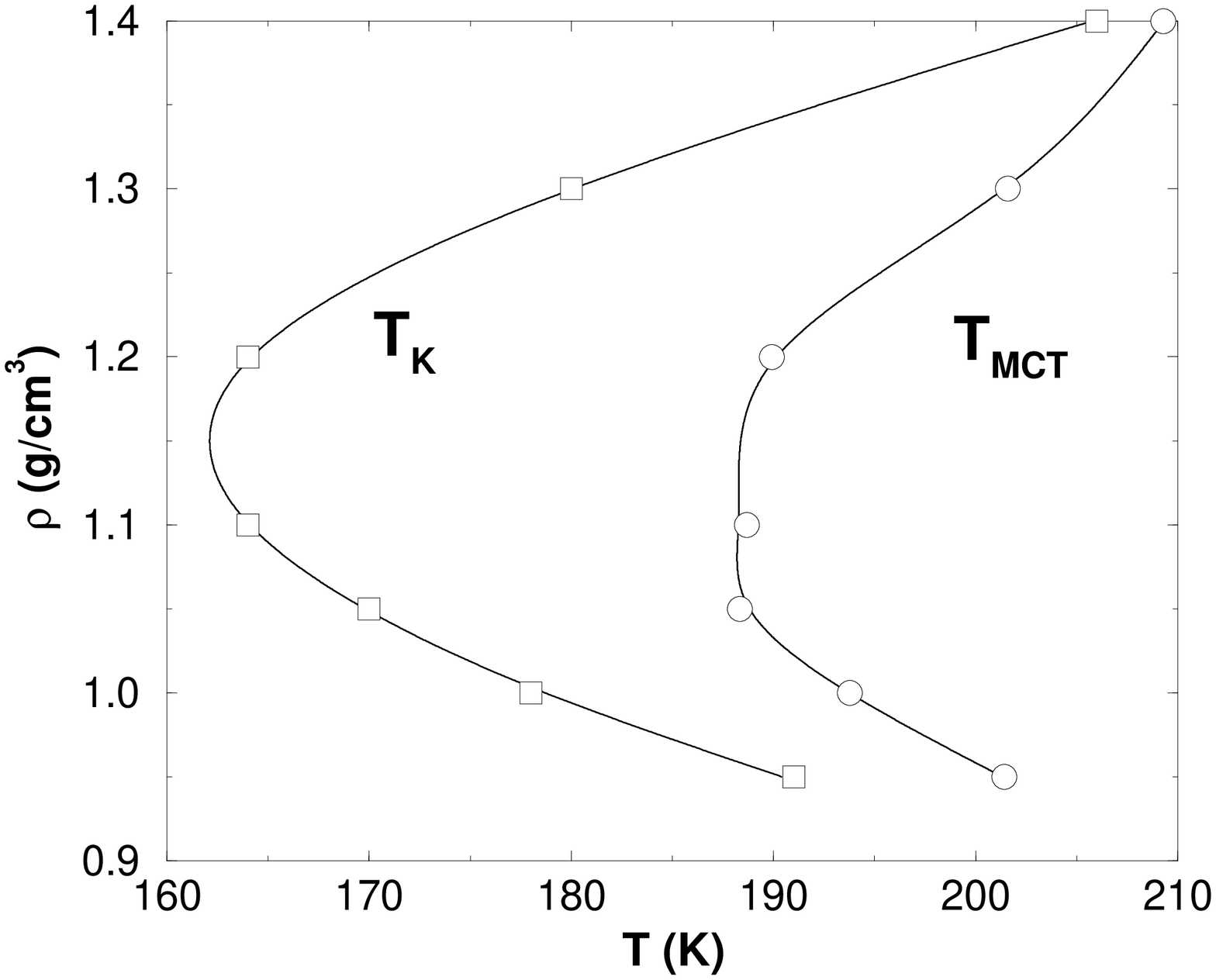,width=8.0cm,angle=-90}}
\caption{The locus of the mode-coupling theory 
transition temperature $T_{MCT}$ (from Ref. \protect\cite{shss}),
and the locus of the Kauzmann temperature $T_K$.}
\label{fig:isoentro}
\end{figure}

Before concluding, we note that the present approach allows us to
estimate the locus of the Kauzmann temperature $T_K(\rho)$ at which
$S_{\mbox{\scriptsize conf}}$ would disappear on extrapolating to lower
$T$.  Fig.~\ref{fig:isoentro} shows the calculated values of $T_K(\rho)$
together with the locus of mode coupling temperatures $T_{MCT}(\rho)$.
The ratio $T_{MCT}/T_K$ has been used as an indication of the fragility
of a liquid~\cite{angell}; we find $T_{MCT}/T_K \approx 1.05 - 1.15$
suggesting that SPC/E is an extremely fragile liquid in the region of
temperatures near $T_{MCT}$ --- as also found in experimental
measurements.  We note that the values of $T_K(\rho)$ depend strongly on
the validity of the extrapolation of the potential energy\cite{T35-note}
to temperatures lower than the one we can equilibrate with the present
computer facilities. Slower changes in $E_{\mbox{\scriptsize conf}}$
below the lowest simulated state point would produce a much slower
decrease for $S_{\mbox{\scriptsize conf}}$~\cite{TKnote}.  This may be
quite plausible since it appears that $E_{\mbox{\scriptsize conf}}$ is
approaching the crystal value, which is expected to always be less than
the liquid value. A much slower change in $S_{\mbox{\scriptsize conf}}$
is expected to be accompanied by a slower decrease of $D$, which may be
related to a possible fragile-to-strong transition in
water~\cite{fragile-to-strong}.

We thank C.A.~Angell, S.~Sastry, and R.J.~Speedy for helpful
discussions, and the NSF for support; FS acknowledges partial support
from MURST (PRIN 98).

\begin{references}
\bibitem{WSSnote} We define ``weakly supercooled states'' as the
temperature range where the diffusion constant decreases by 3 or 4
orders of magnitude compared to its normal liquid value ($\sim
10^{-5}$~cm$^2$/s).  When $T$ approaches the dynamical critical
temperature predicted by the ideal MCT, the description of the slow
dynamical processes offered by the ideal $MCT$ breaks down.
\bibitem{COMPLEXnote} In the context of spin glasses, the
configurational entropy $S_{\mbox{\scriptsize conf}}$ is normally
referred to as the ``complexity'' $\cal I$.
\bibitem{frenkel} At high $T$ and infinite volume, $S$ is given by the
expression for an ideal gas of rigid triatomic molecules.  We determine
$S$ versus $\rho$ via thermodynamic integration along the $T=1000 K$
isotherm starting from the known analytic value in the infinite volume
limit.
\bibitem{CYCLEnote} Specifically, we integrate the relationship $dS = dE/T +
PdV/T$ along both constant $T$ and constant $V$ paths.  Furthermore, we
have tested the reliability of our results for $S$ by checking that
$\oint dS = 0$ along several closed $P-V$ cycles.
\bibitem{simulation-note} All the relevant simulation details, such as
integration step, potential truncation, and temperature control may be
found in ref.~\cite{shss}.
\bibitem{heating-note} We have checked that systems remain confined
in the same basin up to $T \approx 200 K$ when heating the IS from
$T=0$~K in the time scale of 200~ps.  We have evaluated this anharmonic
contribution from the IS of the simulations at $T=210$~K.  However,
using the IS generated from higher temperature configurations does not
significantly affect our results.
\bibitem{TKnote} In the context of estimating $T_K$ from the analysis
of experimental or numerical data, no distinction can be made between a
scenario where $S_{\mbox{\scriptsize conf}}$ indeed goes to zero at 
$T_K$ (with a clear break in the first $T$ derivative) and a scenario where
the extrapolated $S_{\mbox{\scriptsize conf}}$ has a dramatic change in
slope and approaches zero only at a much lower $T$.
\end{references}
\end{document}